\documentclass[article,showpacs,pra]{revtex4}

\usepackage[T1]{fontenc}
\usepackage[latin9]{inputenc}
\usepackage{color}
\usepackage{amsmath}
\usepackage{amssymb}
\usepackage{graphicx}
\usepackage{esint}
\usepackage{bm}
\usepackage{epstopdf}
\usepackage[english]{babel}
\usepackage{appendix}
\usepackage{braket}
\usepackage{tikz}

\usepackage{dsfont}
\usepackage{gensymb}
\usepackage{caption}
\usepackage{float}

\newcommand{\Tr}{\operatorname{Tr}}

\begin{document}

\title{Bell violations with entangled and non-entangled optical fields}

\author{J. Gonzales, P. S\'{a}nchez, V. Avalos, and F. De Zela }

\affiliation {Departamento de Ciencias, Secci\'{o}n F\'{i}sica,
Pontificia Universidad Cat\'{o}lica del Per\'{u}, Apartado 1761,
Lima, Peru.}





\begin{abstract}
We report Bell violations with classical light prepared in both entangled and non-entangled polarization-path, binary states. Our results show that violations of constraints such as the Bell-Clauser-Horn-Shimony-Holt inequality do not necessarily falsify local-realism. Correlations in the realm of classical statistical optics, which are not of the Bell type, may lead to Bell violations.

\end{abstract}


\maketitle

\section{Introduction}

According to a widely-held view in the physics community, local realism has been proved wrong by experimental Bell violations, modulo some remaining loopholes \cite{mermin,valdenebro,ringbauer}. The two most significant of these -- the locality loophole and the detection loophole -- could be jointly closed only recently and after considerable efforts \cite{hensen1,giustina,shalm,hensen2}, thereby reinforcing the idea that one should definitely renounce to local realism. From all Bell violations, those referring to the Bell-Clauser-Horn-Shimony-Holt (BCHSH) inequality \cite{clauserb} might be ranked among the most emblematic ones. Said inequality reads
\begin{equation}\label{eq:1}
    S_{\text{Bell}} \equiv \lvert \eta(a,b)+\eta(a',b)+\eta(a,b')-\eta(a',b') \lvert \leq 2.
\end{equation}
Here, $\eta(a,b)$ stands for the correlation between results obtained by measuring two observables, $A$ and $B$, the settings of which can be chosen between two possible ones in each case, i.e., between $a$, $a^{\prime}$ and $b$, $b^{\prime}$, respectively. Multiple experiments performed with quantum systems have yielded $S_{\text{Bell}}>2$, in accordance with the quantum prediction that $S_{\text{Bell}}\in \left[0,2\sqrt{2}\right]$. This has led to the conclusion that a consistent physical description of natural phenomena cannot be both local and realistic, and that quantum mechanics may be interpreted as being a non-local and/or non-realistic theory \cite{popescu}. However, as we show in this work, $S_{\text{Bell}}\in \left[0,2\sqrt{2}\right]$ can also be achieved within a fully local-realistic framework, such as the one offered by classical light. This means that Bell violations do not necessarily rule out a local-realistic description of natural phenomena. An apparent contradiction with Bell's theorem can arise, if one interprets this theorem as stating that the validity of inequality (\ref{eq:1}) logically follows from the sole assumptions of locality and realism. If that would be the case, there would be no local-realistic experiment that could possibly violate the BCHSH inequality. However, as we shall discuss in more detail below, the derivation of the BCHSH inequality requires that we make some restrictive assumptions regarding the correlations $\eta(a,b)$ entering (\ref{eq:1}). Thus, by considering a different type of correlations, Bell violations become possible within a local-realistic framework. The question then arises, as to which extent quantum Bell violations may be traced back to the difference between quantum correlations and Bell-type correlations, rather than to an alleged non-realistic and/or non-local nature of quantum phenomena. We hope that the results we report in this work will contribute to partially clarify these issues.

The paper is organized as follows. After some theoretical preliminaries, we describe the experiments we have performed and we report our results. The meaning and scope of these results are then discussed, followed by our conclusions.

\section{Theoretical preliminaries}

\begin{figure}[h]
 \centering
   \includegraphics[scale=0.45]{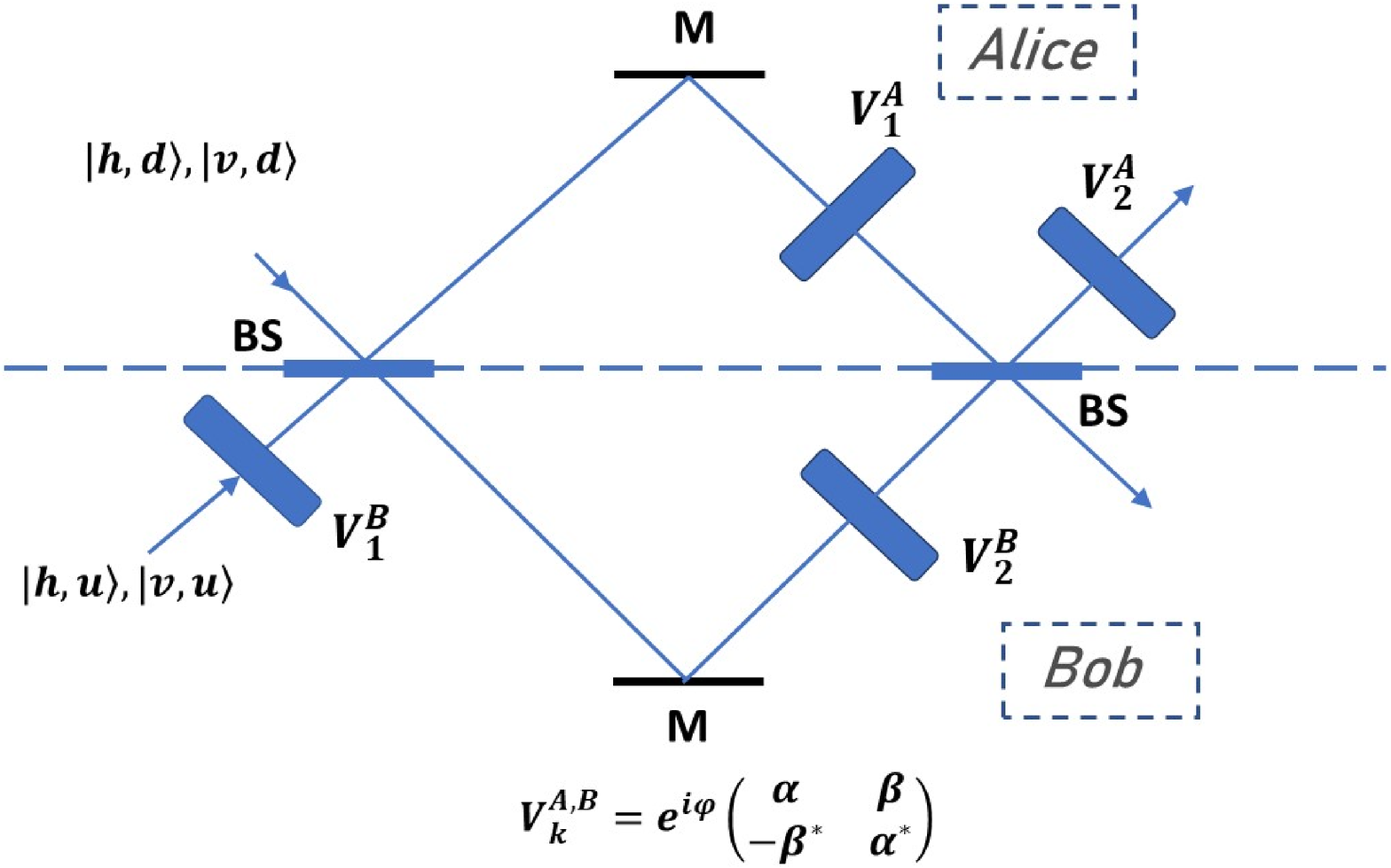}
   \caption{Mach-Zehnder configuration with retarders and phase shifters that perform $U(2)$ transformations $V_{k}^{A,B}$. These can be performed on Alice's and on Bob's sides, independently from one another. From the recorded measurements one assesses correlations such as $\eta_{AB}$ (see text). BS: Beam-splitter, M: mirror.}\label{Fig1}
\end{figure}

Quantum Bell violations are usually obtained with two-qubit, entangled states. With the help of a Mach-Zehnder configuration of the type discussed in Ref. \cite{englert}, we can generate two-qubit states of the form
\begin{equation}\label{col3a}
  |\Phi_{AB}\rangle=r_1 e^{i\phi_1}|h\rangle |x\rangle + r_2 e^{i\phi_2}|h\rangle |y\rangle + r_3 e^{i\phi_3}|v\rangle |x\rangle +r_4 e^{i\phi_4}|v\rangle |y\rangle ,
\end{equation}
with $\sum_{i=1}^4 r_i^{2}=1$ and $|h\rangle$ ($|v\rangle$) denoting a horizontally (vertically) polarized beam, while $|x\rangle$ and $|y\rangle$ stand for any other binary degree of freedom (DOF), in our case a two-way path-DOF. One can in fact implement state $|\Phi_{AB}\rangle$ using optical, classical-light beams. Following \cite{fdz}, we address the reduced density matrices $\rho_A=\Tr_B|\Phi_{AB}\rangle \langle \Phi_{AB}|$ and $\rho_B=\Tr_A|\Phi_{AB}\rangle \langle \Phi_{AB}|$. By writing them in terms of the identity matrix $\sigma_{0}$ and the Pauli matrices $(\boldsymbol{\sigma}=(\sigma_1,\sigma_2,\sigma_3))$ as
\begin{equation}\label{co8}
    \rho_k=\frac{1}{2}\left(\sigma_0+\mathbf{S}_{(k)}\cdot \boldsymbol{\sigma}\right) \equiv \frac{1}{2}\left(\sigma_{0}+\mathcal{P}\mathbf{\hat{n}}_{(k)}\cdot \boldsymbol{\sigma}\right),
\end{equation}
we define the corresponding Stokes vectors $\mathbf{S}_{(k=A,B)}$, whose common modulus, $\mathcal{P}$, is the degree of polarization \cite{wolf}:
\begin{equation}\label{co9}
  \mathcal{P}=\sqrt{S_{(k)1}^{2}+S_{(k)2}^{2}+S_{(k)3}^{2}}.
\end{equation}
The Stokes vectors are thus given by $\mathbf{S}_{(A)}=\Tr_A(\rho_A \boldsymbol{\sigma})$ and $\mathbf{S}_{(B)}=\Tr_B(\rho_B \boldsymbol{\sigma})$, and we may define a correlation between them as
\begin{equation}\label{co14}
    \eta_{AB} \equiv \frac{\mathbf{S}_{(A)} \cdot \mathbf{S}_{(B)}}{\parallel\mathbf{S}_{(A)}\parallel\parallel\mathbf{S}_{(B)}\parallel},
\end{equation}
where $\parallel \cdot \parallel$ means the Euclidean norm. From Eqs. (\ref{col3a}, \ref{co8}, \ref{co14}), with $r_{i=1,\ldots, 4}=1/2$, we get
$\eta_{AB}=\cos(\phi_2-\phi_3)$. In this case, our Bell-parameter reads
\begin{equation}\label{co15}
 S_{\text{Bell}}\equiv \lvert \cos(\phi_2-\phi_3)+\cos(\phi_2-\phi_3^{\prime})+\cos(\phi_2^{\prime}-\phi_3)-\cos(\phi_2^{\prime}-\phi_3^{\prime})\rvert,
\end{equation}
and we can certainly get $S_{\text{Bell}}> 2$, thereby violating inequality (\ref{eq:1}). As we can see, $S_{\text{Bell}}$ depends on four parameters:
$\delta_1 =\phi_2-\phi_3$, $\delta_2 =\phi_2^{\prime}-\phi_3$, $\delta_3 =\phi_2-\phi_3^{\prime}$, $\delta_4 =\phi_2^{\prime}-\phi_3^{\prime}$. These parameters are not independent from one another, because of the identity
$\delta_1+\delta_4 \equiv \delta_2+\delta_3$. If we fix two of the $\delta$'s and use this identity, then we are left with a single free parameter on which $S_{\text{Bell}}$ depends. We have exploited this simplification to perform our experiments. These are variants of those proposed in \cite{fdz}, where it was stressed the parallelism that holds between classical and quantum violations of the BCHSH inequality. Indeed, as illustrated in Fig. (\ref{Fig1}), a Mach-Zehnder-type configuration may be used to prepare bipartite states that span a polarization-path space with basis $\left\{|h,u\rangle, |h,d\rangle, |v,u\rangle, |v,d\rangle \right\}$, where $\ket{h,u} \equiv \ket{h} \ket{u}$, etc. Here, the binary path-DOF has basis vectors $|u\rangle$ and $|d\rangle$, which stand for the up and down propagation directions, respectively (see Fig. (\ref{Fig1})).

We may imagine that there are two parties, Alice and Bob, who submit the DOFs to different transformations by means of wave plates and phase shifters. These transformations can be represented by unitary operators $V^{A,B}_{k} \in U(2)$ of the form
\begin{equation}\label{v}
  V^{A,B}_{k}=e^{i\varphi}\left(\begin{array}{cc}
                            \alpha & \beta \\
                            -\beta^{\ast} & \alpha^{\ast}
                          \end{array}\right),
\end{equation}
where $\alpha$ and $\beta$ are complex numbers such that $|\alpha|^{2}+|\beta|^{2}=1$.
Alice and Bob may choose, independently from one another, how to change the DOFs of the light-beam that is within their reach. To this end, Alice performs the operations $V^{A}_{k=1,2}$, while Bob performs the operations $V^{B}_{k=1,2}$, cf. Fig. (\ref{Fig1}). Thereafter, some measurements can take place, for instance the necessary ones to construct a quantity of interest, e.g., the correlations defined by Eq. (\ref{co14}). This procedure contains all the essential ingredients leading to a Bell violation, similarly to quantum Bell tests in which Alice and Bob perform, e.g., spin measurements with Stern-Gerlach apparatuses, the orientations of which they can choose independently from one another, and being given by unit vectors $\boldsymbol{\hat{a}}$ and $\boldsymbol{\hat{b}}$. The recorded results can then be used to construct quantities such as spin-spin correlations, and with them the parameter
$ S_{\text{Bell}}$ that enters the BCHSH inequality. The formal equivalence between $\eta_{AB}$ of Eq. (\ref{co14}) and quantum correlations having the structure $\eta_{QM} \sim \boldsymbol{\hat{a}}\cdot \boldsymbol{\hat{b}}$ or variants thereof, makes it possible to produce both quantum and classical Bell violations. Notice that we may assume that the apparatuses shown in Fig. (\ref{Fig1}) have been set at positions that are sufficiently far away from each other, so as to guarantee no causal connection between Alice and Bob's settings. The setup of Fig. (\ref{Fig1}) is thus potentially capable of avoiding the signalling loophole. We should however stress that we are not concerned here with issues related to open loopholes in Bell tests, but with their alleged capability to falsify local-realism under ideal conditions (loophole-free tests). Our results preclude concluding that local realism is falsified by violations of the BCHSH inequality, even under ideal conditions. 

\section{Experimental Tests}

We addressed two particular cases of the general state $|\Phi_{AB}\rangle$ and performed experiments with non-entangled states:
\begin{equation}\label{es}
    \ket{\phi} = \frac{1}{2}\left(\ket{h}+\ket{v}) \otimes (\ket{x}+e^{i \delta} \ket{y}\right),
\end{equation}
and with entangled ones:
\begin{equation}\label{nes}
    \ket{\psi}=\frac{1}{\sqrt{2}}\left(\cos\theta\ket{h,x}+\ket{h,y}+\sin\theta\ket{v,x}\right).
\end{equation}
We recall that $\ket{x}$ and $\ket{y}$ represent a binary path-DOF. While in the Mach-Zehnder configuration $\ket{x}$ and $\ket{y}$ represent two different optical paths, in the Sagnac setup the path-DOF is realized by clockwise and counterclockwise light propagation. When dealing with state $ \ket{\phi}$, we used a Sagnac-type setup. In this case, the robust Sagnac configuration allowed us to accurately control $\delta$. In the case of $ \ket{\psi}$, a Mach-Zehnder configuration was stable enough to accurately fix $\theta$. Correlations are given by $\eta_{\phi}(\delta) =\cos\delta$ and $\eta_{\psi}(\theta) =\sin\theta$, respectively. These correlations can lead to Bell violations, as our experiments have shown (see Fig. (\ref{espiral})). All our measurable quantities were intensities. Thus, by carrying out intensity measurements with photodetectors having a linear response to light intensity, we could obtain for states (\ref{es}) and (\ref{nes}) the correlations $\eta_{AB}$ given by equation (\ref{co14}).
In order to characterize one DOF, we performed successive projections over the orthogonal components of the other DOF. Making reference to the general state $|\Phi_{AB}\rangle$ in Eq. (\ref{col3a}), our procedure is based on the following results.

Let us start with $\rho_{AB}=|\Phi_{AB}\rangle \langle \Phi_{AB}|$ and project it over each of the two components of the first DOF. In our case, this is the polarization DOF. The projected states are thus $\rho_{h}^{AB}=P_h \rho_{AB}P_h$, with $P_h=|h\rangle \langle h|\otimes \sigma_{0}^{\text{(path)}}$, and similarly for $\rho_{v}^{AB}$. We can now define $\rho_h^{\text{(path)}}=\Tr_{\text{pol}}(\rho_{h}^{AB})$ and $\rho_v^{\text{(path)}}=\Tr_{\text{pol}}(\rho_{v}^{AB})$, which represent (non-normalized) states in path-space. The diagonal components of $\rho_h^{\text{(path)}}$ and $\rho_v^{\text{(path)}}$ can be experimentally obtained through intensity measurements on each arm ($x$ and $y$) after each projection ($h$ and $v$). Thus, $I_x^{h}=\rho_{h,11}^{\text{(path)}}$, $I_{y}^{h}=\rho_{h,22}^{\text{(path)}}$, $I_x^{v}=\rho_{v,11}^{\text{(path)}}$, $I_y^{v}=\rho_{v,22}^{\text{(path)}}$. We define total intensities associated to horizontal/vertical projections as $I_h^{\text{(path)}}=I_x^{h}+I_y^{h}$ and $I_v^{\text{(path)}}=I_x^{v}+I_y^{v}$, and construct
\begin{equation}\label{path}
  \rho_{\text{path}}=\Tr_{\text{pol}}\rho_{AB} = \left(\frac{I_h^{\text{(path)}}}{I_h^{\text{(path)}}+I_v^{\text{(path)}}}\right)\frac{\rho_h^{\text{(path)}}}{\Tr\rho_h^{\text{(path)}}}+
  \left(\frac{I_v^{\text{(path)}}}{I_h^{\text{(path)}}+I_v^{\text{(path)}}}\right)\frac{\rho_v^{\text{(path)}}}{\Tr\rho_v^{\text{(path)}}}.
\end{equation}
The last equality above is a mathematical identity, as one can readily verify. The Stokes vector that belongs to $\rho_{\text{path}}$ reads therefore:
\begin{equation}\label{spath}
    \mathbf{S}_{\text{path}}=\left(\frac{I_h^{\text{(path)}}}{I_h^{\text{(path)}}+I_v^{\text{(path)}}}\right)\mathbf{S}_{h}^{\text{(path)}}+
    \left(\frac{I_v^{\text{(path)}}}{I_h^{\text{(path)}}+I_v^{\text{(path)}}}\right)\mathbf{S}_{v}^{\text{(path)}}.
\end{equation}
The normalized density matrices $\widehat{\rho}_{h/v}^{\text{(path)}}=\rho_{h/v}^{\text{(path)}}/\Tr\rho_{h/v}^{\text{(path)}}$ are projectors, as one can easily check. Indeed, a straightforward calculation shows that $\widehat{\rho}_{h/v}^{\text{(path)}}\cdot \widehat{\rho}_{h/v}^{\text{(path)}}=\widehat{\rho}_{h/v}^{\text{(path)}}$, so that
$\widehat{\rho}_{h/v}^{\text{(path)}}=|\psi_{h/v}^{\text{(path)}}\rangle \langle \psi_{h/v}^{\text{(path)}}|$, where
$|\psi_{h}^{\text{(path)}}\rangle=a_x|x\rangle+a_y e^{i \alpha}|y\rangle$, and similarly for $|\psi_{v}^{\text{(path)}}\rangle$. While $a_x$ and $a_y$ are already known: $a_x^2=I_x^{h}/(I_x^{h}+I_y^{h})$ and $a_y^2=I_y^{h}/(I_x^{h}+I_y^{h})$, $\alpha$ must be separately measured. This was done by interferometry; that is, again by intensity measurements.

The determination of $\rho_{\text{pol}}$ follows the same pattern, with obvious replacements: projections are done by path selection, while polarization tomography is performed in the standard way, using two retarders and one polarizer in the configuration Quarter-wave-plate/Half-wave-plate/Polarizer (QHP). Because we took care of realizing a $50:50$ beam-splitting, the Stokes vector in the case of the polarization DOF is given by
\begin{equation}\label{spol}
    \mathbf{S}_{\text{pol}}=\frac{1}{2}\mathbf{S}_{x}^{\text{(pol)}}+\frac{1}{2}\mathbf{S}_{y}^{\text{(pol)}}.
\end{equation}

We did not require to displace our optical elements in order to produce relative phase shifts, such as the one entering the path-state $\ket{x}+e^{i \delta} \ket{y}$. Even in this case, we produced the phase-shift $\delta$ with the help of retarders, by exploiting the fact that we deal with path-polarization bipartite states, so that any phase-shift does not distinguish whether it comes from the polarization or from the path DOF.
This and other features allowed us to optimize the setups used for preparation and measurement of states $\ket{\phi}$ and $\ket{\psi}$, as we describe in what follows.
\subsection{Product state}
\begin{figure}[h]
    \centering
    \includegraphics[scale=0.7]{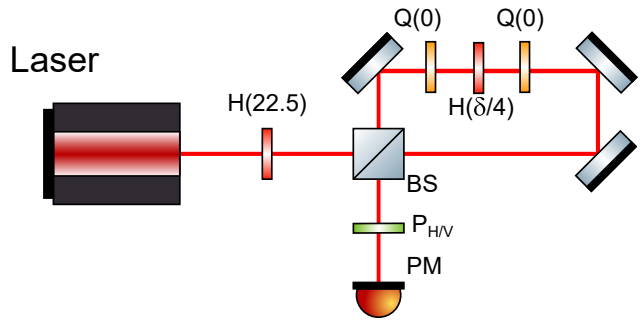}
    \caption{Sagnac-type setup used to prepare state $|\phi \rangle$ and to measure its corresponding path DOF. The polarizer ($\text{P}_{\text{H/V}}$) set before the power meter (PM) selects horizontally or vertically polarized states. Q: quarter-wave plate. H: half-wave plate. BS: beam-splitter.}
    \label{sagnac}
\end{figure}

\begin{figure}[h]
    \centering
    \includegraphics[scale=0.8]{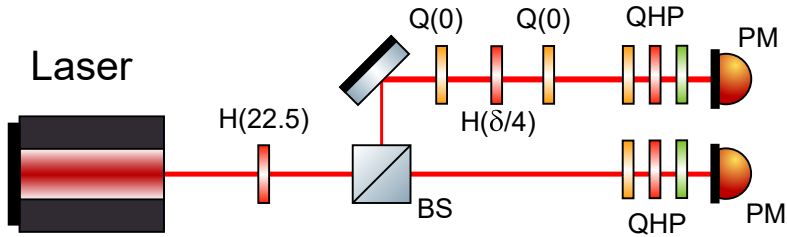}
    \caption{Setup used to measure the polarization DOF of state $|\phi \rangle$. The device $Q(0)H(\delta/4)Q(0)$ plays no role in this case, as it only provides a global phase. Polarization tomography is made with the $QHP$ configuration.}
    \label{sagnac2}
\end{figure}

Figure (\ref{sagnac}) shows the setup used for preparation and measurement of the product state $|\phi \rangle=\left(\ket{h}+\ket{v}) \otimes (\ket{x}+e^{i \delta} \ket{y}\right)/2$. This state can be written in the form $|\phi \rangle=\ket{d}\ket{x} + e^{i \delta}\ket{d}\ket{y}$, with $\ket{d}=(\ket{h}+\ket{v})/2$. To generate $|\phi \rangle$, we first transformed $\ket{h} \rightarrow \ket{d}$ with a half-wave plate set at $22.5^{\circ}$, whereupon we produced (up to normalization) state
$|\phi^{\prime} \rangle=\ket{d}\ket{x} + \ket{d}\ket{y}$ by means of a beam-splitter. On the Sagnac setup following the beam-splitter (see Fig. (\ref{sagnac})) the configuration $Q(0)H(\delta/4)Q(0)$ performs the change $\ket{d} \rightarrow e^{i \delta/2}\ket{d}$ in the beam that goes counterclockwise, and the change $\ket{d} \rightarrow e^{-i \delta/2}\ket{d}$ in the beam going clockwise. Thus, the whole transformation amounts to $|\phi^{\prime}\rangle \rightarrow |\phi \rangle$, up to a global phase. In this case, the Stokes vectors of each beam, $\mathbf{S}_{x}^{\text{(pol)}}$ and $\mathbf{S}_{y}^{\text{(pol)}}$, are just the ones corresponding to the diagonal state $|d\rangle$. We confirmed this by performing standard polarization tomography. To this end, we removed the two mirrors on the right part of the Sagnac configuration and mounted on each of the two resulted arms a three element device ($QHP$) to perform polarization tomography (see figure (\ref{sagnac2})). We did not remove the configuration $Q(0)H(\delta/4)Q(0)$, as it only introduces a global phase without effect on the polarization state.

As for the Stokes vector $\mathbf{S}_{\text{path}}$, its measurement requires measuring $\mathbf{S}_{h/v}^{\text{(path)}}$, cf. Eq. (\ref{spath}). We recall that $\mathbf{S}_{h}^{\text{(path)}}$ belongs to a pure state $|\psi_{h}^{\text{(path)}}\rangle=a_x|x\rangle+a_y e^{i \delta}|y\rangle$, where $a_x^2=I_x^{h}/(I_x^{h}+I_y^{h})$ and $a_y^2=I_y^{h}/(I_x^{h}+I_y^{h})$, while $\delta$ is the phase defined in Eq. (\ref{es}). For the vertically polarized state we have, similarly, $|\psi_{v}^{\text{(path)}}\rangle=b_x|x\rangle+b_y e^{i \delta}|y\rangle$. By appropriately setting a power meter on the Sagnac setup, we could accurately measure the required intensities. As for the phase $\delta$, we measured it by recording the intensity $I_{\text{PM}}$ at the output of the BS with a power meter having a polarizer in front of it to select the corresponding polarization, $h$ or $v$ (see figure (\ref{sagnac})). The two measured intensities,
$I_{\text{PM}}^{h}=I_x^{h}+I_y^{h}+I_x^{h}I_y^{h}\cos(\delta)$ and
$I_{\text{PM}}^{v}=I_x^{v}+I_y^{v}+I_x^{v}I_y^{v}\cos(\delta)$, yielded the same value of $\delta$, within experimental accuracy.

\subsection{Entangled state}

Let us now describe our procedure to prepare and measure state $|\psi \rangle=\left(\cos\theta\ket{h,x}+\ket{h,y}+\sin\theta\ket{v,x}\right)/2$. We observe that this state can be written in the form $|\psi \rangle=\left(\ket{\theta}\ket{x}+\ket{h}\ket{y}\right)/2$, with $\ket{\theta}=\cos\theta\ket{h}+\sin\theta\ket{v}$. Thus, after submitting a horizontally polarized state $\ket{h}$ to a beam-splitter, one needs to perform the change $|h\rangle \rightarrow |\theta \rangle$ on one output beam, leaving the other output beam unchanged. Hence, by setting a half-wave plate at angle $\theta/2$ on the $x$-arm, we transform $\ket{h}\ket{x}+\ket{h}\ket{y} \rightarrow \ket{\theta}\ket{x}+\ket{h}\ket{y}$, thereby preparing state $\ket{\psi}$.

\begin{figure}[h]
    \centering
    \includegraphics[scale=0.8]{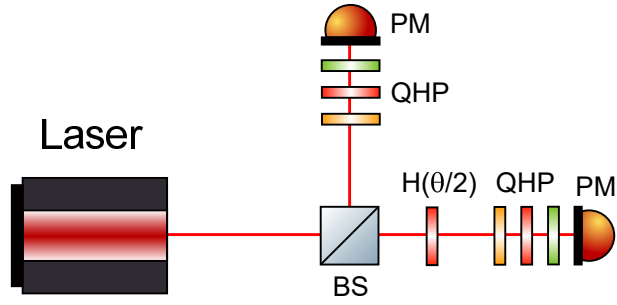}
    \caption{Setup used to prepare state $|\psi \rangle$ and to measure its polarization DOF.}
    \label{poltom}
\end{figure}

\begin{figure}[h]
    \centering
    \includegraphics[scale=0.8]{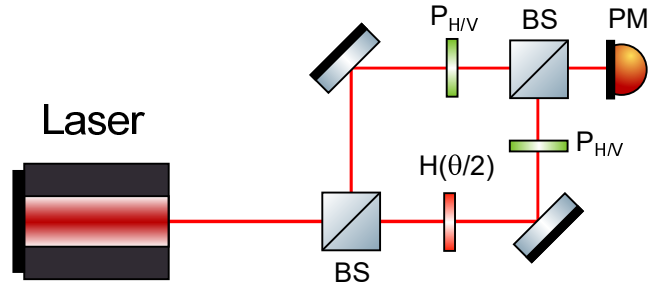}
    \caption{Setup used to prepare state $|\psi \rangle$ and to measure its path DOF.}
    \label{pathtom}
\end{figure}

Stokes vectors $\mathbf{S}_{\text{pol}}$ and $\mathbf{S}_{\text{path}}$ could be obtained as follows. $\mathbf{S}_{\text{pol}}$ required performing polarization tomography on each output beam of the BS, cf. figure (\ref{poltom}), thereby obtaining $\mathbf{S}_{x}^{\text{(pol)}}$ and $\mathbf{S}_{y}^{\text{(pol)}}$. We thus get
$\mathbf{S}_{\text{pol}}=\left(\mathbf{S}_{x}^{\text{(pol)}}+\mathbf{S}_{y}^{\text{(pol)}}\right)/2$, according to equation (\ref{spol}).
To measure Stokes vector $\mathbf{S}_{\text{path}}$, we employed the setup shown in figure (\ref{pathtom}). The procedure is similar to the one used with state $|\phi \rangle$. Here again, Stokes vectors belong to pure states
of the form $|\psi_{h}^{\text{(path)}}\rangle=a_x|x\rangle+a_y e^{i \alpha}|y\rangle$ and
$|\psi_{v}^{\text{(path)}}\rangle=b_x|x\rangle+b_y e^{i \beta}|y\rangle$, with $a_x^2=I_x^{h}/(I_x^{h}+I_y^{h})$ and $a_y^2=I_y^{h}/(I_x^{h}+I_y^{h})$, and similarly for $b_x$ and $b_y$. Due to the form chosen for state $|\psi\rangle$, phases $\alpha$ and $\beta$ are in this case either $0$ or $\pi$. They can be measured in the following way. By setting polarizers that filter, say, horizontal states on each arm of the interferometer shown in figure (\ref{pathtom}), we obtain maximal intensities whenever $\theta <\pi/2$, meaning that $\alpha=0$, whereas for $\theta >\pi/2$ we have minimal intensities and $\alpha=\pi$. The same holds for $\beta$, after filtering vertically polarized states.

\begin{figure}[h]
\begin{tabular}{ccccc}
\includegraphics[scale=0.6]{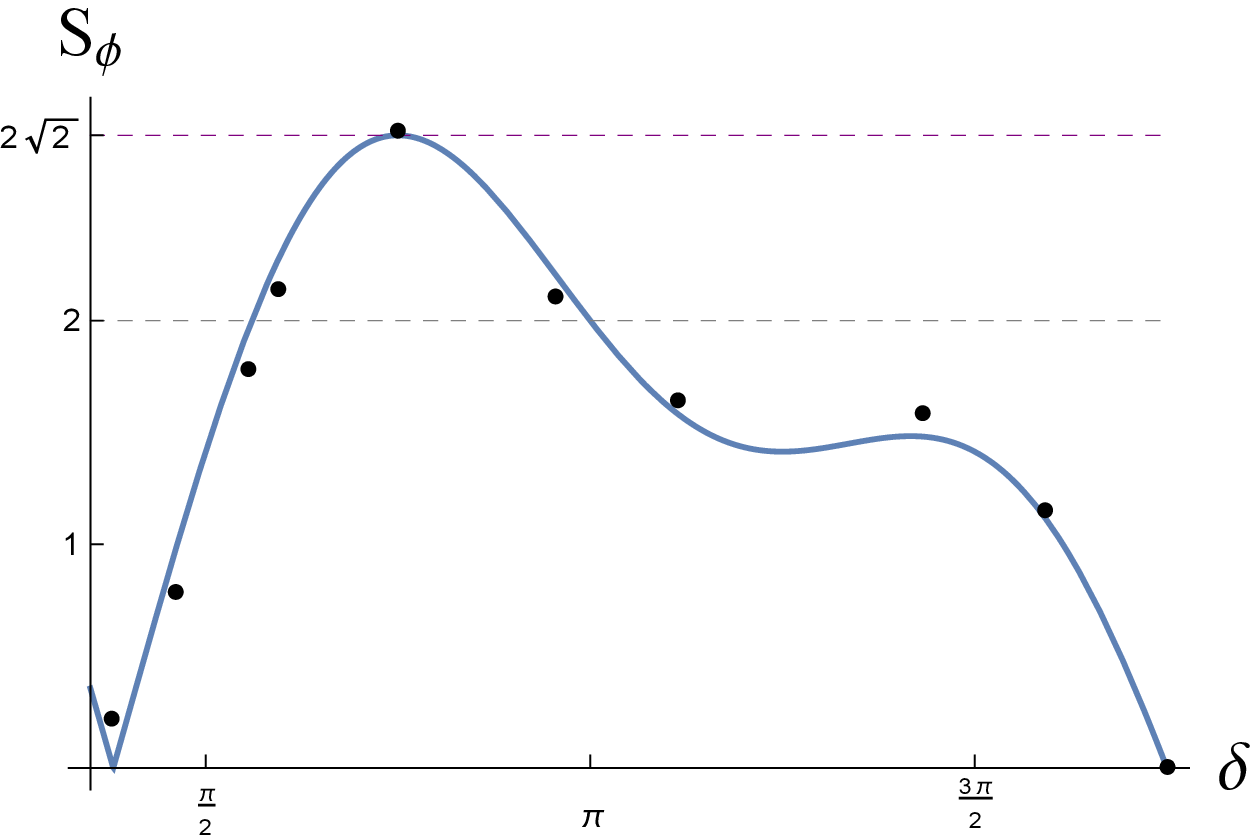}& & & &
\includegraphics[scale=0.6]{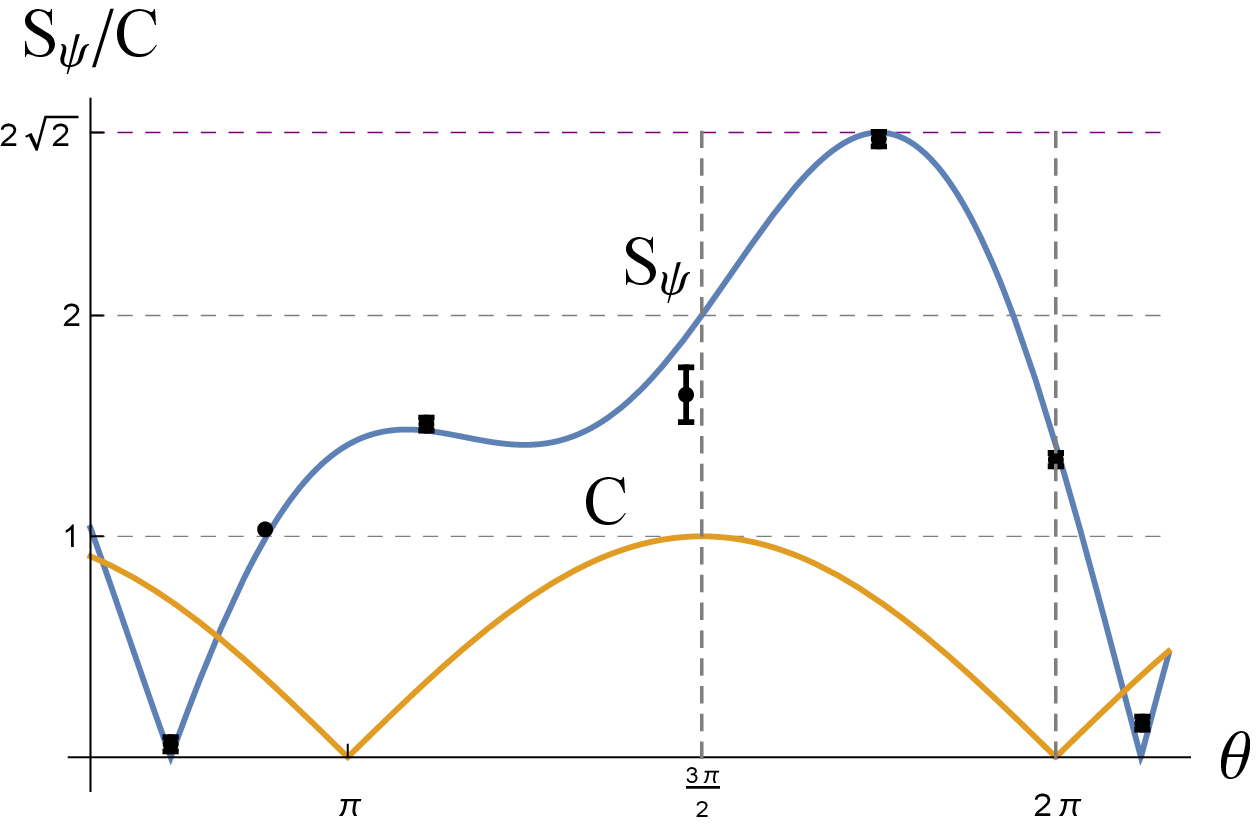}\\
\end{tabular}
\caption{Left panel: Bell parameter $S_{\phi}$ for non-entangled states. Right panel: Bell parameter $S_{\psi}$ for states of variable entanglement. The curve bounded by $0$ and $1$ is concurrence (C), a standard measure of entanglement. Error bars are generally small, similarly sized as symbols, with exception of $\theta \approx 3\pi/2$, for which Stokes vectors are close to zero.}\label{espiral}
\end{figure}

Having obtained $\mathbf{S}_{\text{pol}}$ and $ \mathbf{S}_{\text{path}}$, we calculated correlations $\eta_{\phi}(\delta)$ and $\eta_{\psi}(\theta)$ as per Eq. (\ref{co14}) and constructed Bell parameters $S_{\phi}=|\eta(\delta_1)+\eta(\delta_2)+\eta(\delta_3)-\eta(\delta_4)|$ and $S_{\psi}=|\eta(\theta_1)+\eta(\theta_2)+\eta(\theta_3)-\eta(\theta_4)|$, which are plotted in Fig. (\ref{espiral}). We took $\theta_k =\delta_k$, with $\delta_1=-3\pi/4$, $\delta_2=\delta_3=\delta$, and on view of the aforementioned identity between the $\delta$'s, we set $\delta_4=\delta_2+\delta_3-\delta_1$. 

Our results exhibit Bell violations for the two correlations, $\eta_{\phi}(\delta)$ and $\eta_{\psi}(\theta)$. In both cases the Bell parameter can reach Tsirelson's bound: $2\sqrt{2}$, which is the maximum that can be attained with quantum correlations and with the ones we have used. With some few exceptions, error bars turned out to be smaller than the symbols used in our plots. Figure (\ref{espiral}), right panel, shows $S_{\psi}$ together with concurrence $C \in [0,1]$, a standard measure of entanglement. At first sight, our results seem to be at odds with the relationship that is commonly assumed to hold between entanglement and Bell violations. Indeed, while under full entanglement ($C=1$) Bell's parameter reaches just the classical bound ($S_{\psi}=2$), Tsirelson's bound ($S_{\psi}=2\sqrt{2}$) is reached with a partially entangled state ($C<1$). This is because in the present case the possibility of Bell violations is given by the type of correlations being addressed: $\eta_{AB}$ of Eq. (\ref{co14}), rather than from the type of states being used. These states may be entangled or non-entangled. Correlations $\eta_{AB}$ are mathematically defined as inner products, and this is the technical reason behind the possibility of our Bell violations.

In a broad sense, though, entanglement is nonetheless behind our Bell violations, because $\eta_{AB}$ itself is non-factorable: it cannot be written as a product of two functions, whereby each function depends on a single degree of freedom. This feature may be taken as the very definition of entanglement \cite{schroedinger,qian,eberly1}.

\clearpage

\section{Discussion}
Let us discuss the meaning and scope of our results. To this end, we recall the derivation of the BCHSH inequality (\ref{eq:1}). One starts assuming a local-realistic framework and makes the assumption that -- contrary to what quantum mechanics postulates -- intrinsic randomness does not exist. Hence, any observed, apparent randomness is just a consequence of incomplete knowledge. One may thus assume the existence of ``hidden variables'' $\lambda$ that completely fix any physical state and, therefore, all possible measurement outcomes. In particular, $\lambda$ fully determines the values $A_{a}(\lambda)$ and $B_{b}(\lambda)$ that Alice's and Bob's observables can take on. In this notation, $A_{a}(\lambda)$
means the dichotomic value ($\pm 1$) that is assigned to an up/down result in a Stern-Gerlach-type measurement on Alice's side, who has oriented her Stern-Gerlach device in a way that is specified by label $a$. The meaning of
$B_{b}(\lambda)$ is similar. Thus, if
$A_{a}(\lambda)+A_{a^{\prime}}(\lambda)=0$, then
$A_{a}(\lambda)-A_{a^{\prime}}(\lambda)=\pm 2$, while if
$A_{a}(\lambda)-A_{a^{\prime}}(\lambda)=0$, then
$A_{a}(\lambda)+A_{a^{\prime}}(\lambda)=\pm 2$. Whence,
\begin{equation}
S(\lambda)\equiv
[A_{a}(\lambda)+A_{a^{\prime}}(\lambda)]B_{b}(\lambda)+
[A_{a}(\lambda)-A_{a^{\prime}}(\lambda)]B_{b^{\prime}}(\lambda)=\pm2.
\label{s1}
\end{equation}
The actual values of $\lambda$ are ruled by some probability distribution $\rho_{\lambda}$, which is normalized according to $\int_\Lambda d\rho_{\lambda}=1$. Here, $\Lambda$ is the space of states that are specified by $\lambda$. Let us set $\langle S\rangle=\int_{\Lambda}S(\lambda)d\rho_{\lambda}$. Taking into account that $|\langle S\rangle|\leq \langle |S|\rangle\equiv \int_{\Lambda}|S(\lambda)|d\rho_{\lambda} =2$, we get the BCHSH inequality:
\begin{equation}
S_{\text{Bell}} \equiv |\langle S\rangle|=|\eta(a,b)+\eta(a,b^{\prime})+\eta(a^{\prime},b)-\eta(a^{\prime},b^{\prime})|\leq
2.
\label{bi}
\end{equation}
The quantities $\eta$ entering (\ref{bi}) are given by
\begin{equation}
\eta(a,b)=\int_\Lambda A_a(\lambda)B_b(\lambda) d\rho_{\lambda}. \label{s2}
\end{equation}
The $\eta$'s are therefore a particular type of correlations, Bell-type correlations, for which inequality (\ref{bi}) holds true.

Some remarks are here in order. First of all, by writing $A_{a}(\lambda)=\pm 1$ and $A_{a^{\prime}}(\lambda)=\pm 1$, we have effectively erased all information regarding directions $a$ and $a^{\prime}$, along which the respective measurements were made. Otherwise, we would write $A_{a}(\lambda)=\pm 1_{a}$ and $A_{a^{\prime}}(\lambda)=\pm 1_{a^{\prime}}$. In that case, we should be able to deal with expressions such as $1_{a} \pm 1_{a^{\prime}}$, if we want to proceed further and obtain testable consequences of our assumptions. To be sure, the step that consists in making the replacement $1_{a} \pm 1_{a^{\prime}} \rightarrow 1 \pm 1$ is perfectly valid and was necessary to obtain the BCHSH inequality; but if we want to experimentally test this inequality, then we should take measures to implement the step $1_{a} \pm 1_{a^{\prime}} \rightarrow 1 \pm 1$. This amounts to erasing the distinction between
$1_{a}$ and $1_{a^{\prime}}$, something that does not occur when performing quantum experiments whose goal is to exhibit Bell violations. Indeed, in order to compare quantum correlations with experimental ones, one must keep track of the directions along which measurements take place. This is so because quantum correlations do depend on these directions. On the contrary, the BCHSH inequality holds for correlations that are effectively independent of device orientations which they only nominally refer to. No wonder, such inequality can be violated when using correlations that do depend on device orientations, such as quantum correlations or the ones we have addressed here. These two correlations have a vectorial character. In contrast, Bell-type correlations (\ref{s2}) have a scalar character that stems from identifying $1_{a}$ with $1_{a^{\prime}}$, thereby writing $A_{a}(\lambda)=\pm 1$, $A_{a^{\prime}}(\lambda)=\pm 1$ and so on.

The possibility of classical Bell violations is therefore given whenever we employ correlations having a vectorial structure and use them to construct the parameter $S_{\text{Bell}}$ of inequality (\ref{eq:1}). The inner-product structure of the correlations we have used, cf. Eq. (\ref{co14}), is essentially the same as that of quantum correlations. Indeed, the latter are defined as quantum averages: $ \eta_{QM}^{AB} \equiv \langle \psi_{AB}| (\boldsymbol{\hat{a}} \cdot \boldsymbol{\sigma}^{A})\otimes(\boldsymbol{\hat{b}} \cdot \boldsymbol{\sigma}^{B})|\psi_{AB}\rangle = \Tr\left(\rho_{AB}\Pi_{AB}\right)$, with $\rho_{AB}=|\psi_{AB}\rangle \langle \psi_{AB}|$ and $\Pi_{AB}=(\boldsymbol{\hat{a}} \cdot \boldsymbol{\sigma}^{A})\otimes(\boldsymbol{\hat{b}} \cdot \boldsymbol{\sigma}^{B})$. The bipartite observable $\Pi_{AB}$ depends on unit vectors $\boldsymbol{\hat{a}}$ and $\boldsymbol{\hat{b}}$, which define the single-party observables $\boldsymbol{\hat{a}} \cdot \boldsymbol{\sigma}^{A}$ and
$\boldsymbol{\hat{b}} \cdot \boldsymbol{\sigma}^{B}$. Hence, quantum correlations rest upon the Born rule: $\langle A \rangle_{\rho} =\Tr (\rho A)$, which has in turn the mathematical structure $\Tr \left(A^{\dagger}B \right) \equiv \boldsymbol{\mathcal{A}}^{\ast} \cdot \boldsymbol{\mathcal{B}}$. Here, $\boldsymbol{\mathcal{A}}$ is a vector whose entries are the rows of matrix $A$, written one after the other. $\boldsymbol{\mathcal{A}}$ is thus a so-called ``vectorialization'' of matrix $A$, and similarly $\boldsymbol{\mathcal{B}}$. As has been discussed elsewhere \cite{fdz2}, the Born rule applies in both the quantum and the classical domain. However, regardless of the Born rule, we are generally free to define correlations in the way that proves most convenient for quantifying how much two observables relate to one another, beyond any possible cause-effect connection that might hold between them. There is nothing that prevents us from defining inner-product-type correlations in a classical framework. Quite on the contrary, this type of correlations is rather common place in classical, statistical optics \cite{wolf}.

We can elucidate the physical implications of assignments such as $A_{a}(\lambda)=\pm 1$ and $A_{a^{\prime}}(\lambda)=\pm 1$, by considering the following situation. Let us assume that two unit vectors, $\boldsymbol{\hat{a}}$ and $\boldsymbol{\hat{a}}^{\prime}$, are alternatively produced by some source and then sent to Alice. Likewise, $\boldsymbol{\hat{b}}$ and $\boldsymbol{\hat{b}}^{\prime}$ are sent to Bob. Upon receiving $\boldsymbol{\hat{a}}$, Alice chooses her reference frame such that $\boldsymbol{\hat{x}}\parallel \boldsymbol{\hat{a}} $. With respect to this reference frame, $\boldsymbol{\hat{a}}=(\pm 1,0,0)$. Alice proceeds similarly upon receiving $\boldsymbol{\hat{a}}^{\prime}$, so that $\boldsymbol{\hat{a}}^{\prime}=(\pm 1,0,0)$. By dropping the zeroes, Alice uses the mathematically unsound, but conveniently abridged notation: $\boldsymbol{\hat{a}}=\pm 1$ and $\boldsymbol{\hat{a}}^{\prime}=\pm 1$. Bob does likewise, thereby obtaining $\boldsymbol{\hat{b}}=\pm 1$ and $\boldsymbol{\hat{b}}^{\prime}=\pm 1$. On considering that these are the only possible results, Bell-type inequalities can certainly be derived by following the same steps that lead to, e.g., the BCHSH inequality. One may even incorporate shared randomness, if desired. However, any correlation that we may define in terms of the above assignments is prone to have a very limited scope. No matter which constraint this correlation is subjected to, this constraint would be easily overcome by other correlations that we may define and experimentally realize. This is so because assignments such as $\boldsymbol{\hat{a}}=\pm 1$ and $\boldsymbol{\hat{a}}^{\prime}=\pm 1$ presuppose two different choices of reference frames. This is in fact what is implied in the derivation of the BCHSH inequality, when one sets $A_{a}(\lambda)=\pm 1$ and $A_{a^{\prime}}(\lambda)=\pm 1$ as a representation of dichotomic results that belong to two different orientations of measurement devices. In contrast, quantum correlations are defined with respect to a single reference frame, and the same holds for the classical correlations we have addressed in this work.


\section{Conclusions}
BCHSH violations by themselves do not rule out the possibility of constructing local-realistic models that are in full accordance with physical phenomena (see, e.g., \cite{qian,eberly1,borges,kagalwala,stoklasa,mclaren,sandeau}). This is so because the BCHSH inequality does not derive from the sole assumptions of realism and locality. A third, independent assumption is that the involved correlations are of the Bell type. Not all correlations that may be defined within a local-realistic framework are of the Bell type. The ones we have considered in this work provide an example that exhibits the rather limited scope of the class containing Bell-type correlations. This suggests the need for a critical re-examination of various Bell-type tests.

\section*{Funding Information}
CONCYTEC-FONDECYT (Grant-Nr. 233-2015-2); DGI-PUCP (Grant-Nr. 441).

\end{document}